\documentclass[aps,twocolumn,groupedaddress,showpacs,amssymb,prl,floatfix,preprintnumbers]{revtex4-2}
\usepackage{graphicx,color}
\usepackage{hyperref}
\usepackage{graphicx}             
\usepackage{amsmath}              
\usepackage{multirow}

\newcommand{\as}{$^{75}$As}

\newcommand{\LaFeCoAsO}{${\rm La} {\rm Fe}_{1-x} {\rm Co}_{x} {\rm As} {\rm O}$}

\newcommand{\BaFeCoAs}{${\rm Ba} ({\rm Fe}_{1-x} {\rm Co}_{x})_2 {\rm As}_{2}$}
\newcommand{\BFA}{${\rm Ba} {\rm Fe}_{2} {\rm As}_{2}$}

\newcommand{\TN}      {$T_{N}$}
\newcommand{\TS}      {$T_{S}$}
\newcommand{\Tc}      {$T_{c}$}

\newcommand{\slrrt}{$(T_1T)^{-1}$}
\newcommand{\slrr}{$T_1^{-1}$}
\newcommand{\ssrr}{$T_2^{-1}$}
\newcommand{\ratio}{$R=T_{1ab}^{-1}/T_{1c}^{-1}$}
\newcommand{\ration}{$R=T_{1c}/T_{1ab}$}

\definecolor{orange}{RGB}{255,165,0}
\definecolor{violett}{rgb}{0.5,0.0,0.5}
\definecolor{kblue}{RGB}{0,0,128}
\definecolor{olive}{RGB}{0,139,0}

\begin{document}

\thispagestyle{myheadings}

\title{Mapping out the spin fluctuations in Co-doped LaFeAsO single crystals by NMR}

\author{Piotr Lepucki$^{1}$, Raphael Havemann$^{1}$, Adam P. Dioguardi$^{1}$, Francesco Scaravaggi$^{1}$, Anja U. B. Wolter-Giraud$^{1}$, Rhea Kappenberger$^{1}$, Sai Aswartham$^{1}$, Sabine Wurmehl$^{1,2}$, Bernd B\"uchner$^{1,2}$ and Hans-Joachim Grafe$^{1}$}
\affiliation{$^1$IFW Dresden, Institut f\"ur Festk\"orperforschung, Helmholtzstraße 20, D-01069 Dresden, Germany\\ $^{2}$Institut für Festk\"orper- und Materialphysik und W\"urzburg-Dresden Cluster of Excellence ct.qmat, Technische Universit\"at Dresden, D-01062 Dresden, Germany}

\date{\today}

\begin{abstract}
We determine the phase diagram of LaFe$_{1-x}$Co$_x$AsO single crystals by using nuclear magnetic resonance (NMR). Up to a nominal doping of $x=0.03$, it follows the phase diagram for F-doped polycrystals. Above $x=0.03$, the F-doped samples become superconducting, whereas for Co-doping the structural and magnetic transitions can be observed up to $x=0.042$, and superconductivity occurs only for higher doping levels and with reduced transition temperatures. For dopings up to $x=0.056$, we find evidence for short-range magnetic order. By means of relaxation-rate measurements, we map out the magnetic fluctuations that reveal the interplay of nematicity and magnetism. Above the nematic ordering, the spin fluctuations in LaFe$_{1-x}$Co$_x$AsO are identical to those in Ba(Fe$_{1-x}$Co$_x$)$_2$As$_2$, suggesting that nematicity in LaFeAsO is a result of the fluctuating spin density wave as well.
\end{abstract}

\maketitle

The interplay of nematicity and magnetism is one of the most discussed topics in iron pnictide research. Two main directions have crystallized, (i) spin fluctuations lead to a breaking of the tetragonal symmetry and induce the nematic order below \TS\ \cite{FangPRB2008,XuPRB2008,PaulPRL2011,FernandesPRL2013,LuScience2014,FernandesNatPhys2014,KretzschmarNatPhys2016,YamakawaPRX2016,ChubukovPRX2016}, and (ii) charge density fluctuations on the Fe $d_{xz}$ and $d_{yz}$ orbitals increase with decreasing temperature and lead to an unequal charge density distribution below \TS , i.e. to an orbital order \cite{LeePRL2009,ChenPRB2010,KontaniPRL2010,BaekNatMat2014,OkPRB2018,HongPRL2020}. One way to shed light on this problem is to compare orbital and spin-fluctuations above \TS , as well as orbital order below \TS\ and spin density wave (SDW) order below the magnetic transition temperature, \TN , in different families of iron pnictides. However, the lack of doped single crystals of the LaFeAsO family has prevented a full comparison with other iron pnictides or chalcogenides until recently \cite{HongPRL2020}.

In this Letter, we present NMR results on LaFe$_{1-x}$Co$_x$AsO single crystals, and compare our data to results published for Co-doped \BFA , FeSe, and polycrystalline F-doped LaFeAsO. We find that the phase diagram of LaFe$_{1-x}$Co$_x$AsO agrees well with that of F-doped samples up to a nominal doping of $x=0.03$. Above this level, the F-doped samples become superconducting, while the Co-doped samples still exhibit nematic and SDW order up to $x=0.042$. Superconductivity is suppressed for these doping levels, and sets in only at $x=0.056$. This sample shows remnants of nematic order and local static magnetism at low temperatures, but no bulk coexistence of superconductivity and magnetism. Finally, $x=0.06$ is the only concentration with bulk superconductivity, albeit with a strongly reduced \Tc\ compared to F-doping. Our results show that in-plane doping pins the magnetism in LaFeAsO and suppresses superconductivity. Contrasting the spin fluctuations measured by the spin-lattice relaxation rate, \slrrt , in LaFe$_{1-x}$Co$_x$AsO with those of \BaFeCoAs\ shows that above \TS , \slrrt\ is identical for samples with the same \TS . This indicates that the leading instability for low doping levels is towards a SDW in both compounds, in agreement with recent theoretical predictions~\cite{YamakawaPRX2016,ChubukovPRX2016}.   

Co-doped single crystals of LaFeAsO were prepared by solid state crystal growth as described elsewhere \cite{KappenbergerJCrysGro2018}. The sample sizes were smaller than 1 x 1 x 0.1~mm, and the maximal weight about 0.3~mg. NMR has been measured on the \as\ nucleus for magnetic fields $H$ oriented along [100]$_{ortho}$ (or [010]$_{ortho}$), and along [001] (called $a$, $b$, and $c$). Due to the particular position of As in the crystal structure, \as\ NMR is an excellent probe of stripe-type spin fluctuations in iron pnictides, which can be measured by the spin-lattice relaxation rate and its anisotropy \cite{KitagawaJPSJ2009,SKitagawaPRB2010,NakaiPRB2012,NingPRB2014,GrafePRB2014,DioguardiPRB2015,BohmerPRL2015,ZhouPRB2016,KissikovNatCom2018,ZhangPRB2018,BaekNatQuantMat2020,GrafePRB2020}. Furthermore, the NMR spectra split at the nematic ordering temperature due to an anisotropy of the Knight shift and of the electric quadrupole interaction, and can thus distinguish the $a$ and $b$ directions in the orbital ordered state \cite{FuPRL2012,BaekNatMat2014,BohmerPRL2015,ZhouPRB2016,OkPRB2018,Toyoda1PRB2018,Toyoda2PRB2018,BaekNatQuantMat2020}.

\begin{figure*}
\begin{center}
 \includegraphics[width=2\columnwidth,clip]{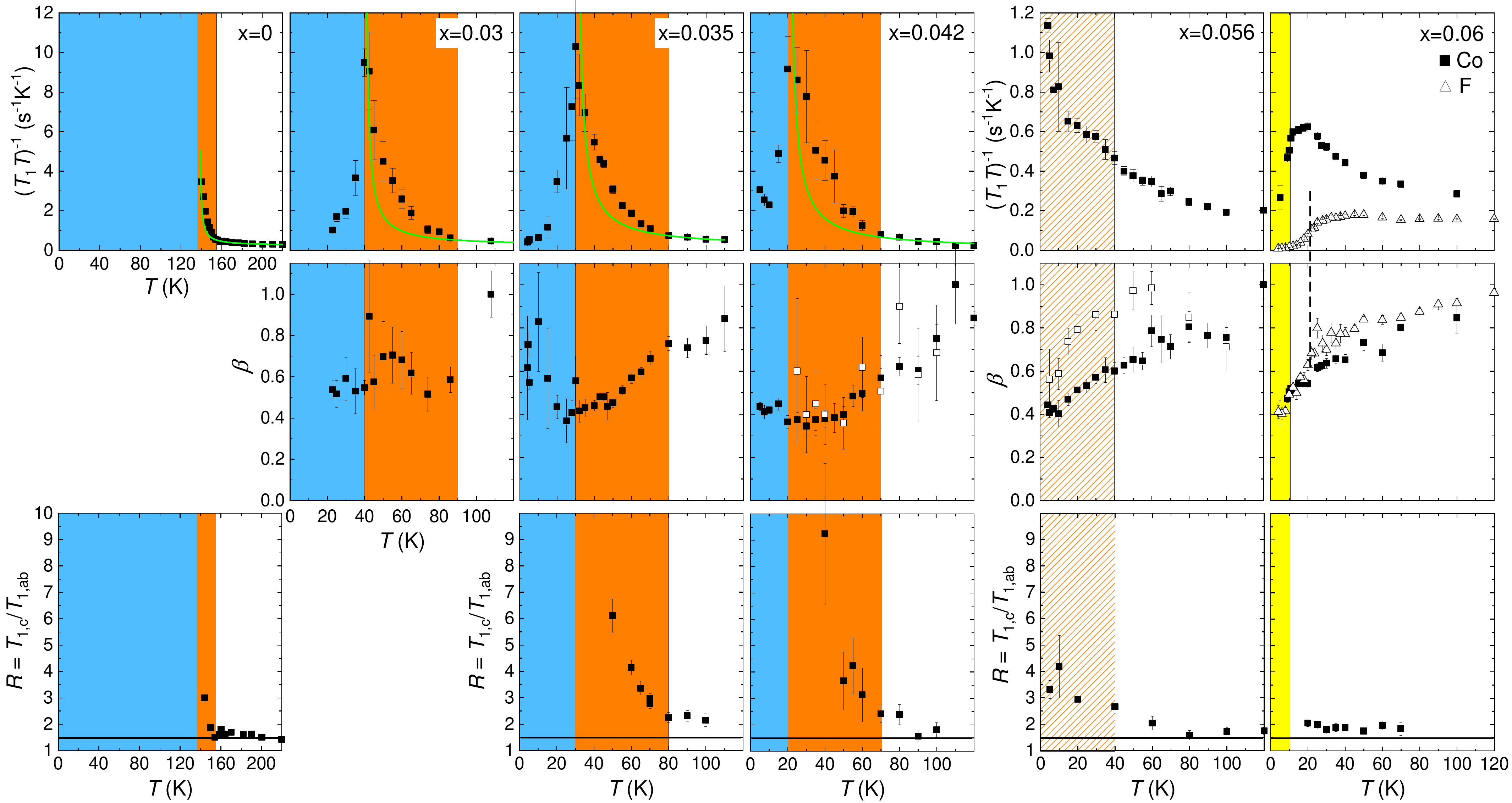}
 \caption{(Color online) Upper row: \slrrt\ versus temperature with the magnetic field applied along the $ab$-plane (parent compound from \cite{OkPRB2018}). In the last panel, \slrrt\ of $x=0.06$ F-doping is included for comparison (dashed line indicates $T_c = 21$~K) \cite{GrafePRB2020}. The scale for $x=0.056$ and $x=0.06$ is 10 times zoomed in compared to the magnetically ordered samples. The green lines are Curie-Weiss fits \cite{footnote1}. Middle row: Stretching parameter $\beta$ of the spin-lattice relaxation rate \slrr, open squares: stretching of the spin-spin relaxation rate \ssrr.  Open triangles: stretching of \slrr\ of $x=0.06$ F-doping. Lower row: Ratio of the spin-lattice relaxation rates, \ration , the solid lines are $R=1.5$. Blue shading indicates bulk antiferromagnetism, orange nematic order, and yellow superconductivity.}
\label{fig:SLRR}
\end{center}
\end{figure*}

The spin-lattice relaxation rate divided by temperature is shown in the upper row of Fig.~\ref{fig:SLRR}. \slrrt\ increases at \TS\ even more rapidly than a Curie-Weiss fit ($(T_1T)^{-1} = a + C/(T-T_N)$) to the data above \TS\ \cite{footnote1}, and reaches a maximum at \TN . This behavior is the same for all samples that order long-range magnetically, i.e. up to $x=0.042$, and indicates a critical slowing of stripe-type spin fluctuations towards the magnetically ordered SDW phase. It is also found in magnetically ordered F-doped polycrystals and in \BaFeCoAs\ single crystals \cite{GrafePRB2020,NingPRB2014,DioguardiPRB2015,KissikovPRB2016}. The maximum of \slrrt\ has been taken as \TN , and the temperature where \slrrt\ increases more rapidly than the Curie-Weiss fit defines \TS . These temperatures agree well with bulk measurements \cite{Scaravaggiunpub,HongPRL2020}. 

For $x=0.056$, \slrrt\ is strongly reduced, and does not reach a maximum anymore, i.e. long-range magnetic order is absent. Note that the scale in Fig.~\ref{fig:SLRR} is zoomed in by a factor of 10 for $x=0.056$ and $0.06$. Critical spin fluctuations above a potential magnetically ordered phase are definitely absent for these two doping levels. For $x=0.056$, \slrrt\ increases monotonically down to the lowest measured temperature. Yet, a change in slope is still visible at a temperature that would agree with a potential structural transition.

Finally, for $x=0.06$, \slrrt\ goes through a broad maximum consistent with a glassy freezing of spin fluctuations at low temperatures without long-range magnetic order \cite{HammerathPRB2013,GrafePRB2020}. Below \Tc , \slrrt\ decreases rapidly due to the opening of the superconducting gap. The slope of \slrr\ below \Tc\ can be described by a power-law with $T^2$, which is smaller than in F-doped samples with $T^3$ or an even higher exponent \cite{HammerathPRB2010,GrafePRB2020}. This is most likely due to the increased disorder by in-plane Co-doping compared to out-of-plane F-doping \cite{Lepuckiunpub}.

The middle row of Fig.~\ref{fig:SLRR} shows the stretching exponent, $\beta$, of the decay of the nuclear magnetization, which indicates a distribution of spin-lattice relaxation rates, where \slrr\ is the median of this distribution \cite{DioguardiPRL2013,JohnstonPRB2006}. Such a distribution is present for all doping levels except for $x=0$. Stretched exponential relaxation often occurs in disordered systems, and is present also for F-doping \cite{HammerathPRB2013,GrafePRB2020}, and in \BaFeCoAs\ for approximately $0.048\leq x\leq0.066$, i.e. for a much smaller doping range \cite{DioguardiPRL2013,NingPRB2014,KissikovPRB2016}. $\beta$ reaches a minimum of about 0.4, indicating that the relaxation rate varies by several orders of magnitude within a sample \cite{JohnstonPRB2006}. More sophisticated analyses of stretched relaxation to inspect the true distribution of \slrr\ \cite{DioguardiPRB2015,SingerPRB2020} could not be performed here owing to the limited signal intensity of the small single crystals.

The lowermost row of Fig.~\ref{fig:SLRR} shows the ratio of \slrrt\ measured for $H||ab$ and $H||c$, \ratio . Due to the peculiar hyperfine coupling of the \as\ nucleus, $R$ is about 1.5 for isotropic spin fluctuations above \TS , and becomes larger than 1.5 below \TS , where the spin fluctuations become anisotropic \cite{KitagawaJPSJ2009,SKitagawaPRB2010,NakaiPRB2012,GrafePRB2014}. All samples that order long-range magnetically also show a clear change of $R$ at the structural phase transition. The change of $R$ for $x=0.056$ is smoothed, indicating that a possible structural phase transition does not occur in the bulk of this sample. Finally, for $x=0.06$, $R$ is constant with temperature, i.e. this sample is tetragonal down to the lowest $T$.

\begin{figure}
\begin{center}
 \includegraphics[width=\columnwidth,clip]{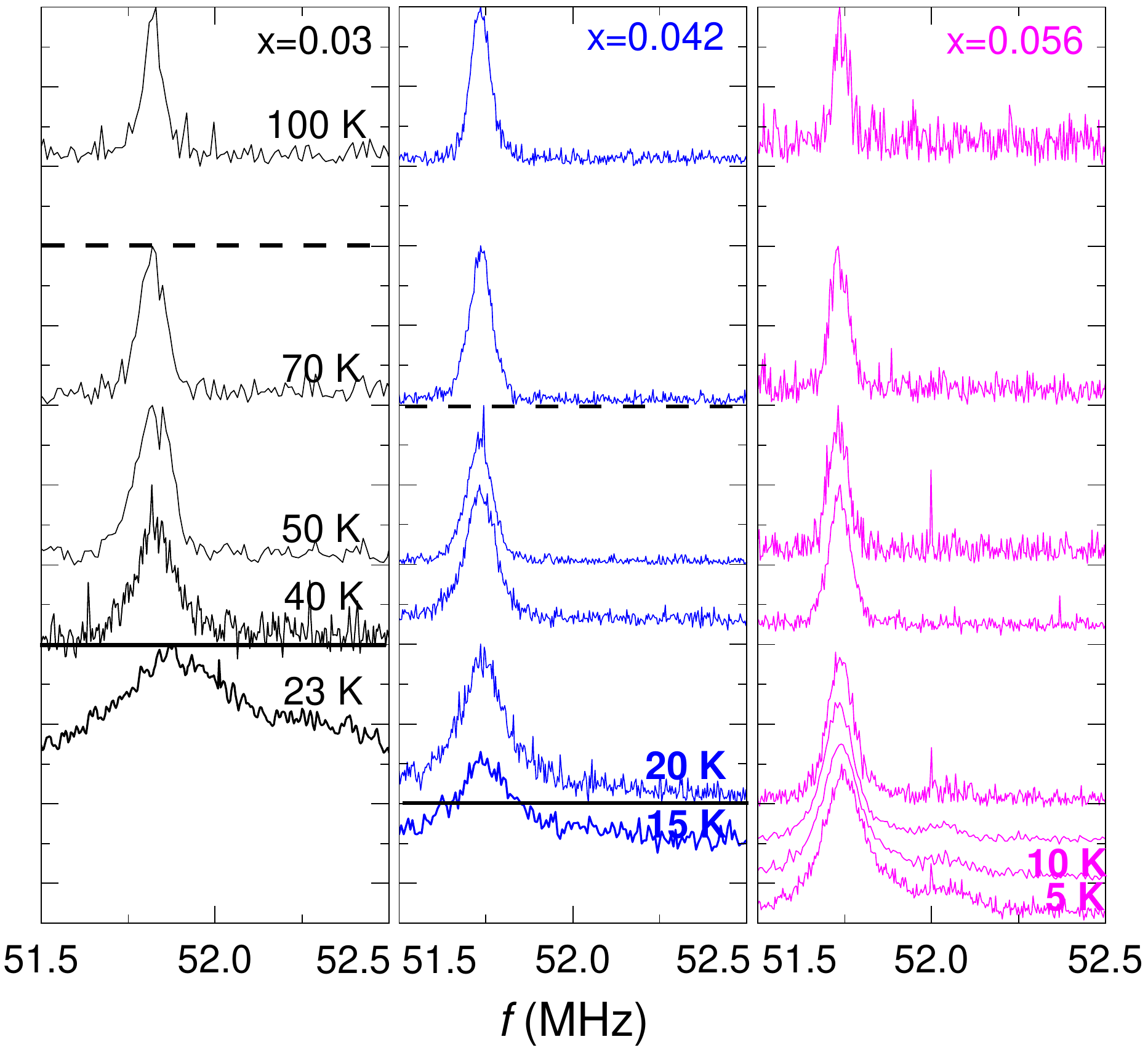}
 \caption{(Color online) NMR spectra for $x=0.03$, 0.042, and 0.056 for $H||[100]_{ortho}$ or $[010]_{ortho}$. Solid lines mark \TN , dashed lines mark \TS . For better visibility, spectra at low temperatures are enlarged.}
\label{fig:spectra}
\end{center}
\end{figure}

\begin{figure}
\begin{center}
 \includegraphics[width=\columnwidth,clip]{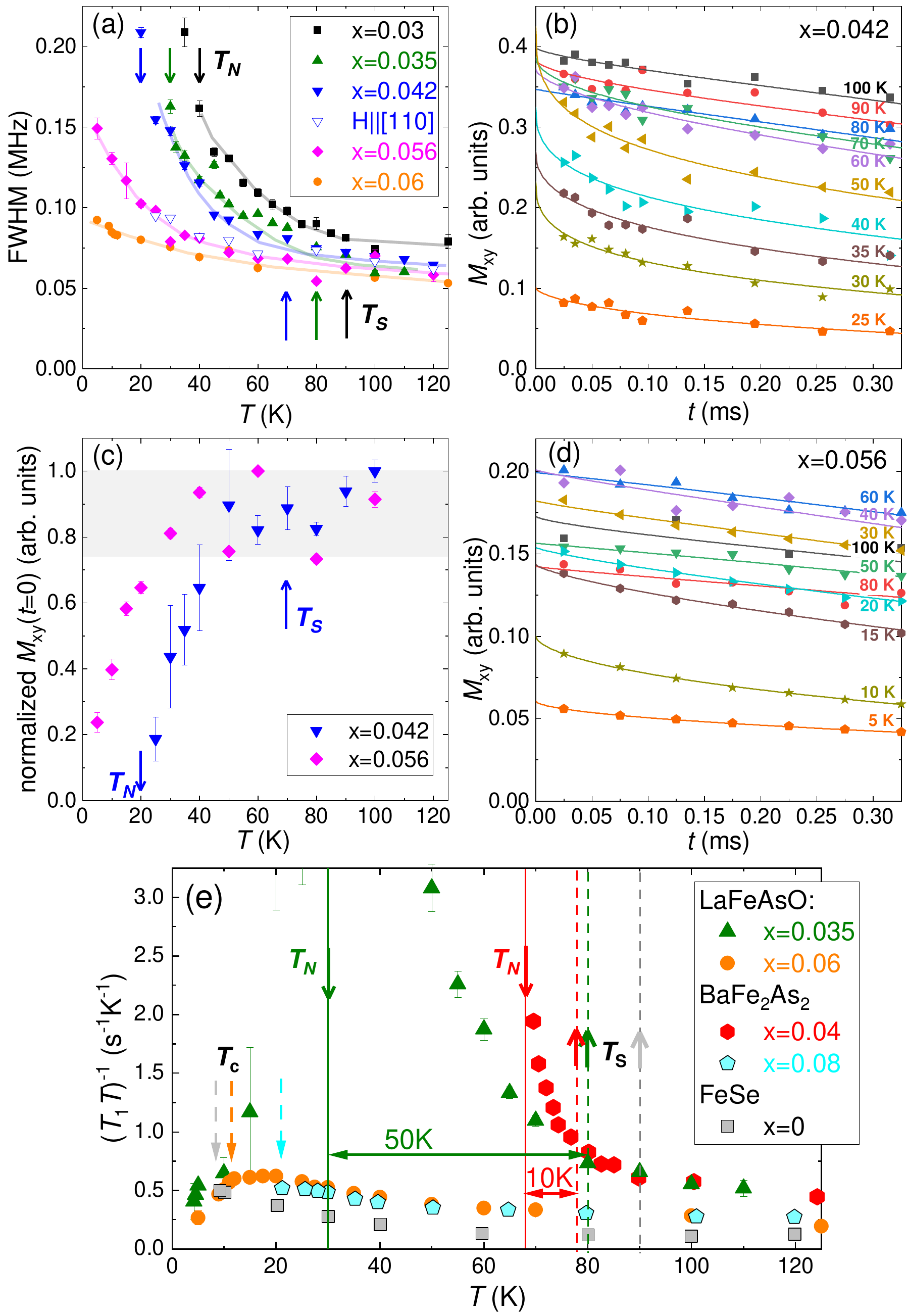}
 \caption{(Color online) (a) FWHM for all doping levels for $H||[100]$ or $[010]$. Open symbols are for $H||[110]$. (b) Spin-spin decay of the nuclear magnetization, $M_{xy}$, for $x=0.042$. (d) Spin-spin decay for $x=0.056$. (c) $M_{xy}$ at $t=0$~s versus temperature for $x=0.042$ and $x=0.056$. (e) \slrrt\ for \LaFeCoAsO\ with $x=0.035$ and $x=0.06$, for \BaFeCoAs\ with $x=0.04$ and $x=0.08$ \cite{NingPRL2010}, and for FeSe \cite{BaekNatMat2014}. Down arrows indicate \TN , dashed-down \Tc , and up arrows \TS .}
\label{fig:T2andwipe}
\end{center}
\end{figure}

The NMR spectra are shown in Fig.~\ref{fig:spectra} exemplarily for $x=0.03$, 0.042, and 0.056 for $H||[100]_{ortho}$ or $[010]_{ortho}$. The spectra for $x=0.035$ and 0.06 are shown in the supplement \cite{supplement}. For all doping levels that exhibit a structural phase transition, the spectra broaden notably below \TS . Surprisingly, the spectra broaden at $T=40$~K for $x=0.056$, too, as can be seen in Fig.~\ref{fig:T2andwipe}(a), where the full width at half maximum (FWHM) is plotted against temperature. The spectra of $x=0.03$ even show remnants of a splitting, which indicates the anisotropy between the $a$ and $b$ directions due to the nematic transition \cite{FuPRL2012,OkPRB2018}. For higher doping levels, the reduced orthorhombicity, $\epsilon = \Delta L/L$,  \cite{WangJMMM2019} prevents resolving the splitting. Undoped LaFeAsO exhibits a splitting of about 75~kHz, which is already on the order of the FWHM for $x=0.03$. The steep increase of the linewidth at \TS\ is another determination method of the structural transition temperatures, which agree well with those determined by \slrrt . Note that for field orientation $H||[110]_{ortho}$, the FWHM increases only moderately with decreasing temperature, see Fig.~\ref{fig:T2andwipe}(a). For this orientation, the spectra of undoped LaFeAsO do not split below \TS\ \cite{OkPRB2018}. The angular dependence of the linewidth below \TS\ can be used to confirm the correct orientation of the samples \cite{supplement}.

Below the SDW transition, the spectra shift to higher frequency and further broaden, indicating the existence of internal static hyperfine fields at the \as\ nuclei. The shift and broadening is typical for the field orientation $H||ab$ \cite{KitagawaJPSJ2009}. With increasing doping, the hyperfine field quickly decreases \cite{NingPRB2014}. The shift and broadening is therefore best visible for the lowest doping level of $x=0.03$ in Fig.~\ref{fig:spectra}. In addition, the spectra develop a shoulder at even higher frequency, indicating that some nuclei feel higher internal magnetic fields. Interestingly, such a shoulder is also visible for $x=0.056$ below about 20~K, and indicates that at least parts of this sample develop static magnetism as well.

Another feature of the spectra is a loss of signal intensity about 15-20~K above the magnetic ordering temperature. Such a wipeout of signal intensity is also present in \BaFeCoAs , and has been ascribed to a shortening of the spin-lattice and spin-spin relaxation rates and their distribution over a few orders of magnitude \cite{DioguardiPRL2013,DioguardiPRB2015}. This is known as a dynamic wipeout, where the nuclei relax so fast that they cannot be observed anymore. Our \slrrt\ and $\beta$ values for $x=0.03$, 0.035, and 0.042 are very close to those observed in underdoped \BFA , and therefore, a shortening of relaxation times seems to be the main reason for wipeout in underdoped LaFe$_{1-x}$Co$_x$AsO. However, the intensity decreases for $x=0.056$ below about 20~K, too, despite the fact that \slrrt\ is about 10 times smaller for this sample.

To further investigate the development of magnetism in the nematic phase, we measured the spin-spin relaxation rate for $x=0.042$ and $x=0.056$, i.e. at the border between long-range magnetic order and superconductivity. The decay of the in-plane nuclear magnetization, $M_{xy}$, is plotted in Fig.~\ref{fig:T2andwipe}(b) and (d). $M_{xy}$ was fit globally at all temperatures for a given sample to a combined exponential and Gaussian form: $M_{xy}(t) = M_0 \cdot \exp(-(2t/T_{2})^\beta) \cdot \exp(-(2t)^2/2T_{2G}^2)$, where $M_0$ is the initial nuclear magnetization, $t$ the time between the $90^\circ$ and $180^\circ$ pulses. $T_2$ is the exponential spin-spin relaxation time, and $T_{2G}$ the Gaussian relaxation time, which turned out not to be temperature dependent \cite{BossoniPRB2013,DioguardiPRB2015,GrafePRB2020,supplement}. 

$\beta \leq 1$ accounts for a distribution of spin-spin relaxation times similar to the distribution of the spin-lattice relaxation times. Yet, in contrast to $T_1$, $\beta$ affects mostly the first few tens of microseconds of the $T_2$-relaxation, which is the time slot that is most difficult to measure due to ringing of the resonance circuit, especially for such small samples as have been measured here \cite{PelcPRB2017}. $\beta$ of $T_2$ is shown in the middle row of Fig.~\ref{fig:SLRR} in comparison to $\beta$ of $T_1$. For $x=0.042$, both $\beta$ show approximately the same temperature dependence, whereas for $x=0.056$, $\beta(T_1)$ is notably smaller than $\beta(T_2)$. However, both data sets in Fig.~\ref{fig:T2andwipe}(b) and (d) clearly show that, even considering a stretching, $M_{xy}(t=0)$ decreases with decreasing temperature (see Fig.~\ref{fig:T2andwipe}(c)). This is evidence for the presence of local static magnetic fields below about $T_{SRO} = 42$~K for $x=0.042$ and below about $T_{SRO} = 23$~K for $x=0.056$, in contrast to the mostly dynamic wipeout observed in \BaFeCoAs . Static short range magnetism has also been found in F-doped LaFeAsO polycrystals below \TS , and has been attributed to a nanoscale phase separation in doped LaFeAsO~\cite{GrafePRB2020}, which we found for Co-doped polycrystals as well \cite{Lepuckiunpub}.

Comparing the low-frequency spin fluctuations above \TS\ in \LaFeCoAsO\ with those in \BaFeCoAs\ indicates whether the nematic order is driven by magnetic fluctuations or by orbital order \cite{FernandesPRL2013,ChubukovPRX2016}. Fig.~\ref{fig:T2andwipe}(e) shows \slrrt\ for two samples for each of the different families, which have similar structural transition temperatures, namely for \BaFeCoAs\ with $x=0.04$ from Ref.~\onlinecite{NingPRB2014} and for \LaFeCoAsO\ with $x=0.035$ from this work. The relaxation rates for these two samples are identical above \TS . Below \TS , \slrrt\ increases much faster with decreasing temperature for \BaFeCoAs . This is also true for the undoped samples except for a constant offset, if the temperature is normalized by \TS\ \cite{supplement}. Differences in spin fluctuations therefore appear only below \TS . The reason for this is possibly the stronger coupling of nematicity and magnetism in \BFA\ compared to LaFeAsO, in agreement with the smaller difference of \TS\ and \TN\ in \BFA . For optimal doping levels which do not exhibit a structural transition, \slrrt\ is again identical for both compounds down to the superconducting transition temperature \Tc\ (see Fig.~\ref{fig:T2andwipe}(e)). In contrast, \slrrt\ of FeSe is even smaller than that of optimally doped LaFeAsO and \BFA , even below \TS\ \cite{footnote2}. Here, the coupling of nematicity and magnetism is even weaker, so that no strong enhancement of low-frequency spin fluctuations appears, neither above nor below \TS , and no static magnetism develops despite of the nematic order at $T_S = 90$~K. These results suggest that the leading instability for LaFeAsO and \BFA\ is the SDW, whereas FeSe shows orbital order without magnetic fluctuations \cite{BaekNatMat2014}, in agreement with recent theoretical considerations~\cite{ChubukovPRX2016,YamakawaPRX2016}.

\begin{figure}
\begin{center}
 \includegraphics[width=\columnwidth,clip]{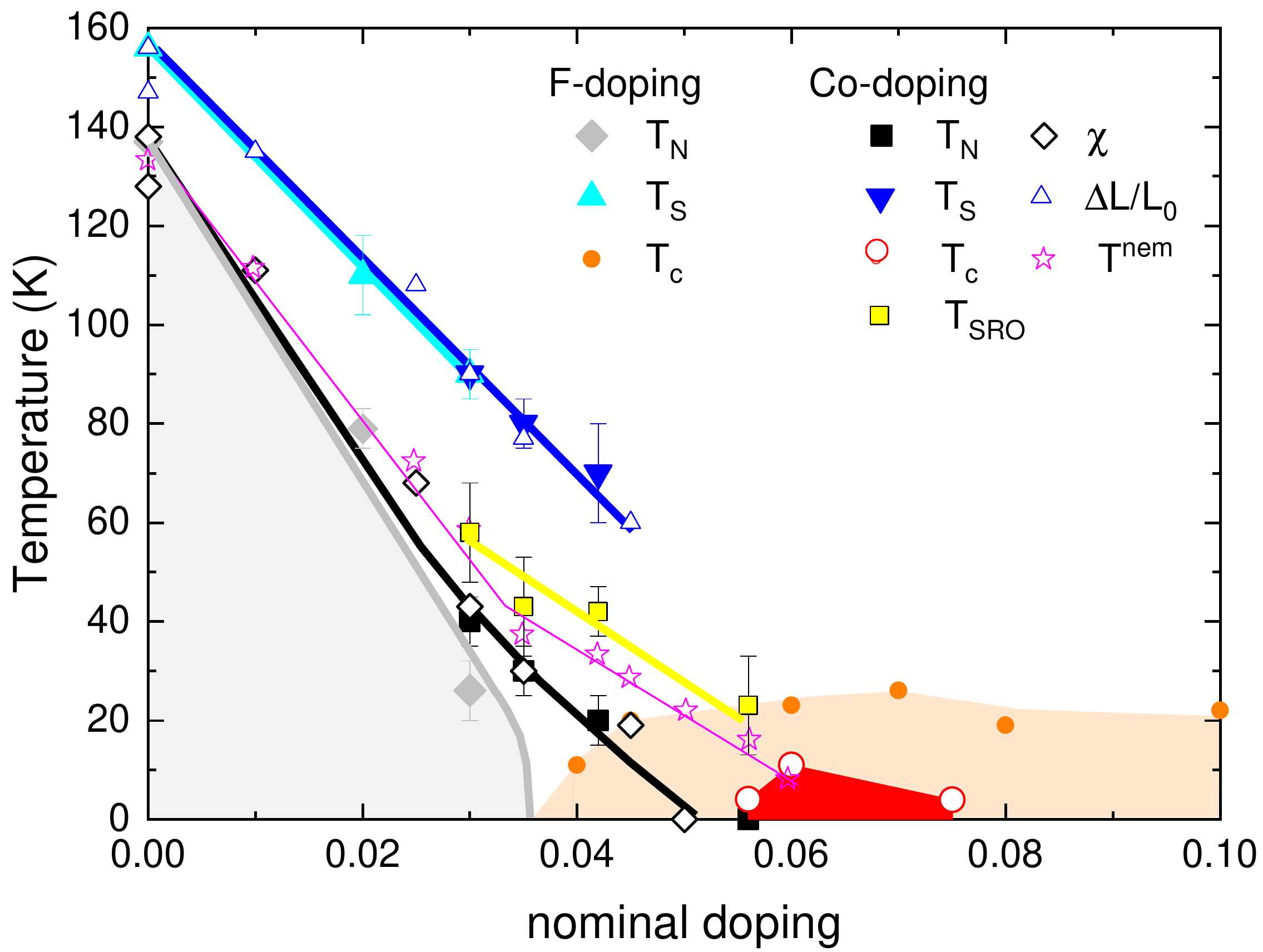}
 \caption{(Color online) Phase diagram resulting from NMR and SQUID susceptibility on Co-doped LaFeAsO single crystals compared to F-doped polycrystals~\cite{GrafePRB2020}. Closed symbols are from NMR, open symbols are from SQUID and dilatometry~\cite{Scaravaggiunpub}. $T^{nem}$ is determined by elastoresistivity \cite{HongPRL2020}.} 
\label{fig:phasedia}
\end{center}
\end{figure}

All results can be combined in the phase diagram in Fig.~\ref{fig:phasedia}. Up to $x=0.03$ the phase diagram of Co-doped LaFeAsO coincides with that of F-doped polycrystals, and \TS\ and \TN\ decrease almost linearly with doping. Above $x=0.03$, F-doped samples become superconducting, and nematicity and magnetism disappear abruptly, whereas for Co-doping, \TS\ and \TN\ can still be detected up to $x=0.042$. This indicates that in-plane Co-doping pins the magnetism to some extend. The sample with $x=0.056$ shows remnants of a structural transition (see Fig.~\ref{fig:SLRR} and \ref{fig:T2andwipe}(a)) as well as short-range magnetic order below $T_{SRO} \approx 23$~K. This is also the first doping that exhibits superconductivity, albeit with a strongly reduced volume fraction. Finally, $x=0.06$ is the only doping with bulk superconductivity and absence of static nematicity and magnetism \cite{Scaravaggiunpub}. Nevertheless, compared to F-doping, the superconducting transition temperature is greatly reduced, and the weaker slope of \slrr\ below \Tc\ indicates scattering at impurities, which are most likely the in-plane Co-dopants. Interestingly, the nematic transition temperature, $T^{nem}$, determined by elastoresistivity measurements~\cite{HongPRL2020} follows the short-range magnetic order temperature determined here, leaving the possibility open that nematicity is related to magnetism in Co-doped LaFeAsO as well.

Compared to Co-doped \BFA , \TN\ and \TS\ decrease much faster with doping in LaFeAsO, which is surprising considering the much larger superconducting range for \BFA\ and the significantly higher \Tc 's. This could be a consequence of the nanoscale phase separation in LaFeAsO~\cite{GrafePRB2020}, where doping leads to a percolation of the magnetic square lattice, and therefore to a more effective suppression of nematicity and magnetism. In contrast, superconductivity and magnetism can coexist in Co-doped \BFA , and no phase separation occurs. However, the normal state spin fluctuations in both compounds are very similar, indicating that nematicity is a result of the fluctuating SDW order in LaFeAsO, as has been predicted by theory recently \cite{ChubukovPRX2016,YamakawaPRX2016}.

We would like to thank Rüdiger Klingeler, Christian Hess and Xiaochen Hong for helpful discussions.
This work has been supported by the Deutsche Forschungsgemeinschaft (DFG) through Grant No. AS 523/4-1, Grant No. DI2538/1-1, research training group GRK1621, through the SFB 1143 (project id 247310070), and the W\"urzburg-Dresden Cluster of Excellence on Complexity and Topology in Quantum Matter ct.qmat (EXC 2147, project id 390858490).


\begin{thebibliography}{80}
\bibitem{footnote1}
There are more data points above 120~K which have been used for the Curie-Weiss fits. Also $\Theta$ has been fixed to \TN .
\bibitem{footnote2}
In FeSe, \slrrt\ has been measured in a magnetic field of 9~T on the $^{77}$Se nucleus, which has a slightly different gyromagnetic ratio. This leads to a reduction of \slrrt\ by at most 15~\% compared to our $^{75}$As NMR on doped LaFeAsO in 7~T. This alone can not explain the 100~\% smaller \slrrt\ in FeSe compared to optimally doped LaFeAsO.
\end{thebibliography}
\end{document}